\documentclass{article}
\usepackage{ismir,amsmath,cite,url}
\usepackage{graphicx}
\usepackage{color}
\usepackage{amsfonts}
\usepackage{tabularx}
\usepackage{array}

\title{Polyphonic Music Generation with Sequence Generative Adversarial Networks}

\oneauthor
{Sang-gil Lee, Uiwon Hwang, Seonwoo Min, and Sungroh Yoon}
{Electrical and Computer Engineering, Seoul National University, Seoul, Korea \\ {\tt \{tkdrlf9202, uiwon.hwang, mswzeus, sryoon\}@snu.ac.kr}}

\begin{document}

\maketitle
\begin{abstract}
We propose an application of sequence generative adversarial networks (SeqGAN), which are generative adversarial networks for discrete sequence generation, for creating polyphonic musical sequences. Instead of a monophonic melody generation suggested in the original work, we present an efficient representation of a polyphony MIDI file that simultaneously captures chords and melodies with dynamic timings. The proposed method condenses duration, octaves, and keys of both melodies and chords into a single word vector representation, and recurrent neural networks learn to predict distributions of sequences from the embedded musical word space. We experiment with the original method and the least squares method to the discriminator, which is known to stabilize the training of GANs. The network can create sequences that are musically coherent and shows an improved quantitative and qualitative measures. We also report that careful optimization of reinforcement learning signals of the model is crucial for general application of the model.
\end{abstract}

\section{Introduction}\label{sec:introduction}

% music generation 이란
Automatic music generation is a concept of creation of a continuous audio signal or a discrete symbolic sequence that represents musical structure from computational models in an autonomous way \cite{hiller1959experimental}. A continuous audio signal includes raw waveform and a spectrogram as a data structure. A discrete symbolic sequence includes MIDI and a piano roll. In this paper, we focus on the polyphonic music generation with MIDI, where the system creates both chords and melodies simultaneously.

% DL짱짱맨
Recent advancements in deep learning \cite{lecun2015deep} have brought a wide range of applications, such as image \cite{he2016deep} and speech recognition \cite{amodei2016deep}, machine translation \cite{cho2014learning}, and bioinformatics \cite{min2017deep}. They are also getting attention for music generation and there have been various approaches \cite{briot2017deep}. Specially, recurrent neural networks (RNNs) are widely used for music language modeling, since they can process time series information which has a central role in musical structure. 

% SeqGAN 이란?
Generative adversarial networks (GANs) \cite{goodfellow2014generative} are frameworks in deep learning that are achieving state-of-the-art performance in generative tasks. However, GANs are more difficult to train with discrete sequences than with continuous data, which results in their limited applications in domains with discrete data. Sequence generative adversarial networks (SeqGAN) \cite{yu2017seqgan} are one of the first models that try to overcome this limitation by combining reinforcement learning and GANs for learning from discrete sequence data. The SeqGAN model consists of RNNs as a sequence generator and convolutional neural networks (CNNs) as a discriminator that identifies whether a given sequence is real or fake. SeqGAN successfully learns from artificial and real-world discrete data and can be used in language modeling and monophonic music generation.

% why polyphonic
The results from the original work have shown a strong potential for application of SeqGANs to automatic music generation. However, the original work have shown rather simple approaches to melody generation (i.e. monophonic music generation) by only using the melody part of the MIDI music and constraining available words in the model to 88-key pitches. In contrast, polyphonic music generation \cite{goel2014polyphonic,johnson2017generating,hadjeres2016deepbach}, where the system can compose both chords and melodies simultaneously, is more appealing and can greatly improve the realism of the computer-generated music.

This consideration leads us to a question of how to represent the language of symbolic music that the model can effectively leverage. We would like to design a word representation of the polyphonic symbolic music with minimal hand-designed preprocessing that would negatively impact the representational power. In addition, we would like to let the model fully incorporate the structure of the data distribution of polyphonic music, including chords, keys, and dynamic timings.

% SeqGAN을 polyphinic music generation 에 써봤음
Based on the pioneering work, we apply SeqGAN for the purpose of polyphonic music generation. Specifically, we propose a simple and efficient word token formulation of polyphonic MIDI sequence that can be learned by SeqGAN. Our representation can capture multiple keys and durations of MIDI music sequence with word embedding. Since we integrated the duration of notes to word representations, the recurrent networks can learn sequences with dynamic timings. The proposed method condenses duration, octaves, and keys of both melodies and chords into a single word vector representation and recurrent neural networks learn to predict distributions of sequences from the embedded musical word space. Sampled sequences from the trained networks show long-term structures that are musically coherent and show an improved quantitative measure of BLEU score and perceptive quality from Mean Opinion Score (MOS) by adversarial training. We discuss about advantages and limitations of the approach and future works.

\begin{figure*} [t]
 \centerline{\framebox{
 \includegraphics[width=\textwidth]{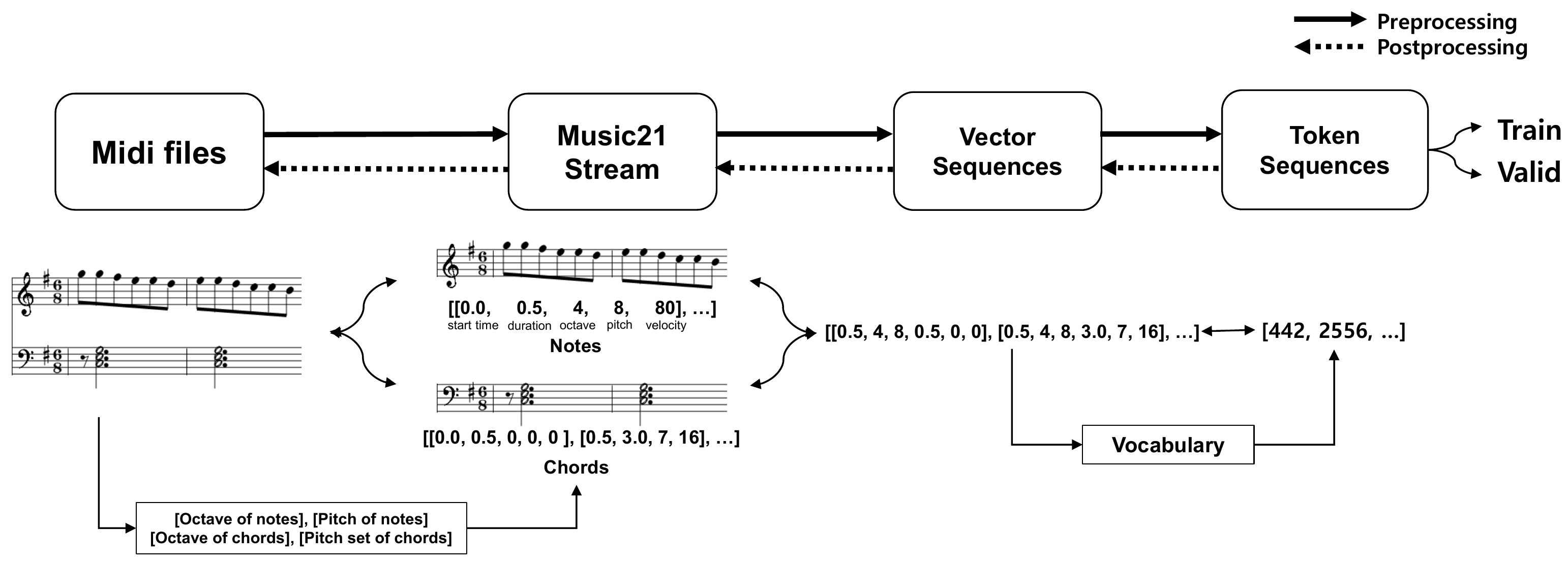}}}
 \caption{Preprocessing and postprocessing pipeline of MIDI files for polyphonic music sequence.}
 \label{fig:data_pipeline}
\end{figure*}

\section{Related Work}
% 미리 작성했던 CNN-RNN과 유사 + SeqGAN을 main related work으로
% survey paper, RNN & LSTM
Refer to \cite{briot2017deep} for a comprehensive survey on deep learning-based music generation. RNNs are widely used for the task of sequence generation, and are designed for processing time-series sequences. Primarily used in language modeling, RNNs can also be applied to music generation based on discrete sequences, notably MIDI and piano rolls. Long Short-Term Memory (LSTM) is a variant module for RNNs that incorporates contextual memory cells and gates for information flow that ``\textit{learn to forget}'' and alleviates the long-term dependency problem of RNNs \cite{hochreiter1997long}. Recent models with RNNs typically use LSTM as a building block.

%% exposure bias of LSTM
Based on the success of the LSTM that can handle long-term dependency, there have been studies of music generation using LSTM. However, there is a problem called ``exposure bias'' \cite{ranzato2015sequence} in the discrete sequence generation using LSTM, when a model is trained with the maximum likelihood method. In the case of an out-of-sample discrete sequence not in the training set, a discrepancy between training and inference occurs because the sampled output of the previous time step is used as the input in the current time step.

%% SeqGAN
SeqGAN \cite{yu2017seqgan} addresses this problem by considering the sequence generation problem as a sequential decision-making process in the reinforcement learning (RL). Further, to calculate reward signals at each time step for RL, SeqGAN incorporates GANs, where the discriminator CNNs provide scores that identify whether the given sequence is real or fake. After being pretrained with a negative log-likelihood (NLL) loss, the generator RNNs are trained by the policy gradient method \cite{sutton2000policy} with these RL signals. More specifically, the generator uses the average of discriminator outputs for sequences generated by Monte Carlo search with a rollout policy as the estimated reward. The rollout policy is set to be the same as the current generator. The generator is updated by the following equations:
\begin{equation}
\begin{split}
\nabla_\theta J(\theta) &=  \mathbb{E}_{Y_{1:t-1}\sim G_\theta} \Big[ \textstyle \sum_{y_t \in \mathcal{Y}} \nabla_\theta G_\theta (y_t|Y_{1:t-1}) \\
 &\cdot Q^{G_\theta}_{D_\phi} (Y_{1:t-1}, y_t) \Big]  \\
&\simeq \frac{1}{T} \sum_{t=1}^{T} \mathbb{E}_{y_t \sim G_\theta (y_t|Y_{1:t-1})} \Big[ \nabla_\theta \log G_\theta (y_t|Y_{1:t-1}) \\
&\cdot Q^{G_\theta}_{D_\phi} (Y_{1:t-1}, y_t) \Big]
\end{split}
\end{equation}
where $G_\theta$ is the policy parameterized by the generator and $Q^{G_\theta}_{D_\phi}$ is the action value function of a sequence following policy $G_\theta$. In an actual implementation, $Q^{G_\theta}_{D_\phi}$ is replaced with the output of the discriminator as mentioned above. $Y_{1:t-1}$ denotes a sequence from the generator and $y_t$ is a token at time step $t$. The parameters of the generator are updated by the gradient ascent method. The parameters of the discriminator are trained with the GAN loss. More detailed explanations can be found in the original SeqGAN paper.

% future work에서 다시 언급할거라서 RL signal 이용한 work 필수
There are other RL approaches in addition to SeqGAN. Using RL for our task has an advantage of the ability to utilize well-defined music theories to calculate reward signals that can be leveraged by the model \cite{jaques2017tuning}. Compared to end-to-end training approaches, RL has an advantage of allowing to guide the network with our prior knowledge of music and steering the model with user preferred musical styles.

% preprocessing에서 다시 언급할거라서 한 음 여러번 치는거 구별못하는 work 필수
The interaction between a composer and the generator is one of the important factors in the music generation task. Therefore, various conditional mechanisms for the music generation have been developed \cite{yang2017midinet,chu2016song}. MidiNet \cite{yang2017midinet} is a model that generates a monophonic note sequence conditioned on a primer melody or a chord sequence. However, symbolic representation of music is not able to distinguish between a single long note and multiple repeating notes in this work.

MidiNet can generate polyphonic music only by priming a given chord as a condition. Our work instead explores the unconditioned polyphonic music generation by distilling all the necessary information into a word embedding space and letting the model to learn from the embedded space. Note that the conditional generation is also possible with our method by priming pre-defined word sequences before the unconditional generation.

C-RNN-GAN \cite{mogren2016c} uses RNNs as a sequence generator and incorporates GANs framework in parallel to our work. However, it uses real-valued feature representation of a MIDI file by modeling tone length, frequency, intensity, and time with four real-valued scalars. RNNs are trained from the real-valued feature space, because of the challenge of training GANs with discrete data, as it was discussed above. Our work is based on the framework that can natively handle a discrete sequence with GANs.

% performance RNN 얘기 - 우리는 정면으로 polyphonic music generation에 도전했고 good performance 획득
Efficient representation of musical data is crucial for the ability of the model to learn the musical structure. Notable examples include Performance RNN\cite{performance-rnn-2017}, which emphasizes that the training dataset and musical representation are the most interesting elements of deep learning-based music generation. Performance RNN uses MIDI representation that handles expressive timing and dynamics, which can be considered as a compressed version of a fixed time step.

\section{MIDI Data Representation}
% nottingham dataset 설명
For our MIDI music dataset we used Nottingham database, which is a collection of 1,200 British and American folk tunes. Note that the original work also used the same dataset, but it only used the monophonic melody part with fixed time steps for training and evaluation. We extend the representation of the dataset for polyphonic sequences. 

% preprocessing & postprocessing 자세한 파이프라인 서술
% 과거형 서술로 바꿨습니다
%% Preprocessing: Parsing
We used music21 Python package for preprocessing of the MIDI data into an input sequence and for postprocessing the output sequence back to MIDI as depicted in \figref{fig:data_pipeline}. A MIDI file in the Nottingham dataset consists of two parts: the melody and chords. After each MIDI file in the dataset was loaded, each note in the file was parsed into a list containing start time, duration, octave, pitch and velocity. For chords, we assigned different indices to all different sets of pitches. For example, [C,E,G] and [G,B,D] have different indices in the pitch list. In this way, we incorporated approximately 30 pitch sets into the pitch list. The statistics of the pitch sets is shown in \tabref{count_chord}. In experiments, we omitted the velocity for two reasons: to reduce the vocabulary size to a tractable amount, and because the incorporation of the velocity would scatter the word distribution severely, which would not yield good estimation results given the amount of data points in the Nottingham dataset. 

\begin{figure*} [t]
 \centerline{\framebox{
 \includegraphics[width=\textwidth]{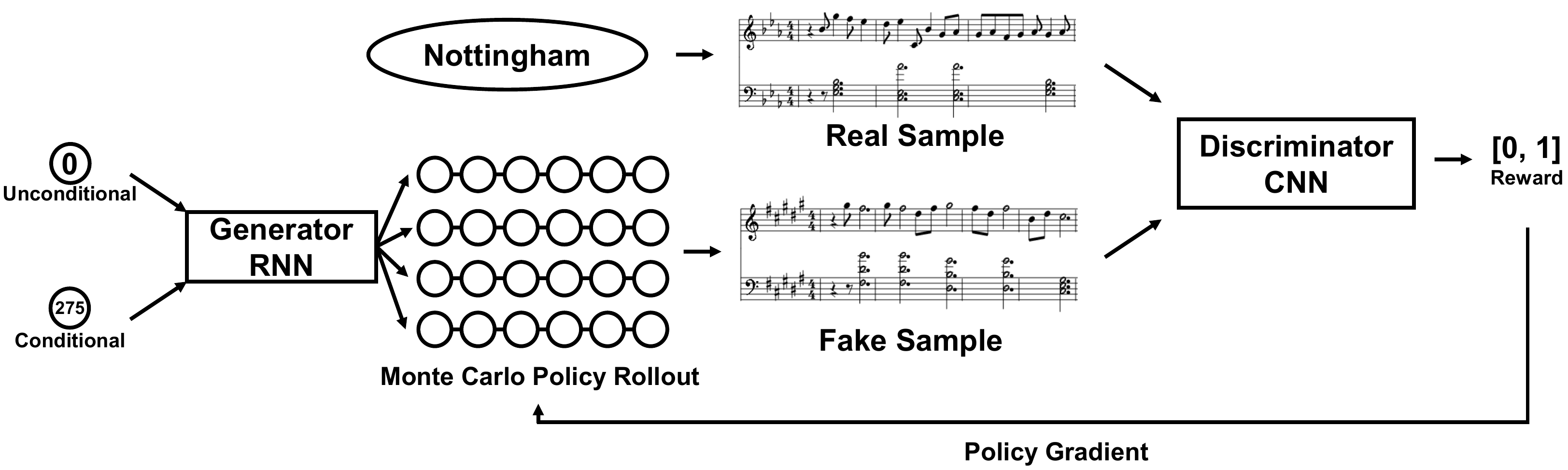}}}
 \caption{Schematic diagram of sequence generative adversarial networks (SeqGAN).}
 \label{fig:model_description}
\end{figure*}

% 의원이 table
% 오른쪽 column 삐져나옴, X로 wrapping함
\def\tabularxcolumn#1{m{#1}}
\newcolumntype{Z}{>{\centering\let\newline\\\arraybackslash\hspace{0pt}}X}
\begin{table}
\begin{center}
 \begin{tabularx}{\columnwidth}{>{\centering}m{2cm}|>{\centering}m{2.7cm}|Z}
  Counts & Pitch sets & Chord symbols\\
  \hline\hline
  \small{5000 $\sim$ 10000} & 
  \footnotesize{[D,G,B], [D,F$^\sharp$,A]} & 
  \small{G/D, D}\\
  \hline
  \small{2000 $\sim$ 4999} & 
  \footnotesize{[C$^\sharp$,E,A], [C,E,G], [E,G,B], [C,E,A]} &
  \small{C$\sharp$m$\sharp$5, C,\newline Em, C6}\\
  \hline
  \small{1000 $\sim$ 1999} & 
  \footnotesize{[C$^\sharp$,E,G,A], [C,D,F$^\sharp$,A]} &
  \small{A7/C$\sharp$, D7/C}\\
  \hline
  \small{500 $\sim$ 999} & 
  \footnotesize{[D,F$^\sharp$,B], [C,F,A], [D,E,G$^\sharp$,B]} &
  \small{D6, F/C, \newline E7/D}\\
  \hline
  \small{250 $\sim$ 499} &
  \footnotesize{[D,G,A$^\sharp$], [D,F,A], [D,F,A$^\sharp$], [E,G$^\sharp$,B]} &
  \small{Gm/D, Dm, \newline A$\sharp$/D, E}\\
  \hline
  \small{100 $\sim$ 249} & 
  \footnotesize{[D,F,G,B], [C$^\sharp$,F$^\sharp$,A], [D$^\sharp$,F$^\sharp$,A,B], [C,E,G,A$^\sharp$], [C,D$^\sharp$,G]} &
  \small{G7/D, F$\sharp$m/C$\sharp$, \newline B7/D$\sharp$, \newline C7, Cm}\\
  \hline
  \small{10 $\sim$ 99} & 
  \footnotesize{[D$^\sharp$,G,A$^\sharp$], [C,D$^\sharp$,F,A], [C$^\sharp$,E,F$^\sharp$,A$^\sharp$], [D,F$^\sharp$], [D$^\sharp$,F$^\sharp$,B], [C$^\sharp$,E,G], [C$^\sharp$,E,G$^\sharp$], [C$^\sharp$,F$^\sharp$,A$^\sharp$], [D,E,G,B]} &
  \small{D$\sharp$, F7/C, \newline F$\sharp$7/C$\sharp$, D, \newline D$\sharp$m$\sharp$5, C$\sharp$dim, \newline C$\sharp$m, F$\sharp$/C$\sharp$, \newline G6/D}\\
\end{tabularx}
\end{center}
\caption{Pitch set statistics of Nottingham dataset.}
\label{count_chord}
\end{table}

%% Tokenize
Tokenization was done by scattering every possible combinations of the musical information into separate words. That is, the duration, the octave of the note, the pitch of the note, the octave of the chord and the pitch of the chord of a time step were combined in a single integer. By including durations in the preprocessing pipeline we were able to tokenize each time step with different lengths. For notes whose lengths were different from the corresponding chords, we inserted dummy notes so that the length of a note and that of a chord sequence would be the same. Rest and dummy notes were designated as a special `rest' token. We excluded music with tokens which occurred less than 10 times in the total dataset to keep the size of the vocabulary tractable. Tokenized integer sequences were used as inputs for SeqGAN.

%% Postprocessing
Based on the generated output sequence of tokens from the SeqGAN model, postprocessing was performed to convert the sequences to MIDI files. After loading the constructed vocabulary with a token sequence, each token in the sequence was converted to two musical symbols, a note and a chord, through the reverse process of the preprocessing. The symbols were appended to the melody stream and the chord stream. After processing all tokens, the two streams were combined into a MIDI file.

%% 장점: 다른모델과 달리 연속으로 같은음치는게 구분됨
% RNN이 fixed time-step 이상을 학습할수 있다는 장점 추가
Unlike in models with fixed time steps introduced in the related work \cite{yang2017midinet}, our preprocessing method can distinguish between a case where a single note is played for a long time and a case where a single note is played multiple times. Our method can do so, because we represented a variable duration by a single word token that can be processed by the recurrent networks. The dynamic timing of this representation can also benefit the generative model, where the RNNs can learn the time-dependent structure of the musical sequence beyond the fixed time steps.

%% 단점: naive hashing으로 input space가 너무 큼
The proposed preprocessing method is designed with minimal human-designed reformulation possible, since we wanted to let the model fully observe the underlying data distribution of polyphonic symbolic MIDI data that the model could leverage from learning. However, our method also has a drawback due to the tokenizing with naive hashing-like approach. Naive hashing can make vocabulary space expand more than necessary. It is difficult to learn chords in an octave that appear only few times in the dataset, even if the same chords in other octaves are abundant in the dataset. For example, tonic triad in different octaves are actually related, but the vocabulary maps to different tokens.

\begin{figure*} [t]
 \centerline{\framebox{
 \includegraphics[width=\textwidth]{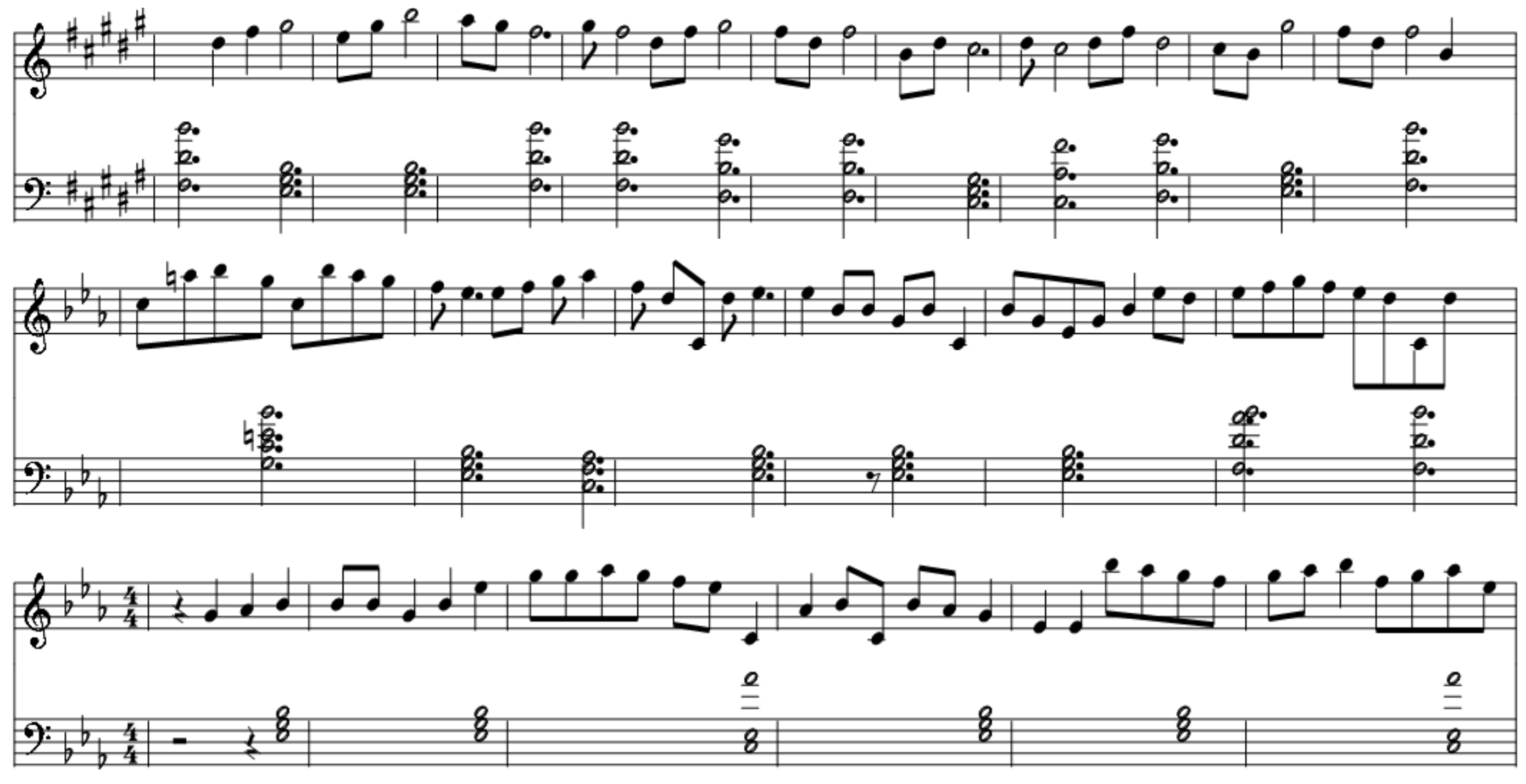}}}
 \caption{Sample music sequences generated from the model.}
 \label{fig:sample_sequences}
\end{figure*}

\section{Model Description}
Here we describe core details of the SeqGAN model and our modifications to the stabilized training of the model with our customized polyphonic MIDI dataset. In SeqGAN, the generator RNNs and discriminator CNNs are pretrained with a regular negative log-likelihood (NLL) loss (until convergence). Then they are further tuned by adversarial training with policy gradient with outputs from the discriminator CNNs ranging from 0 to 1 as reward signals. We followed the same training scheme as in the original work.

% 오리지널 hyperparam은 not working & RL 파라미터 중요
We experienced instabilities in the adversarial training with hyperparameters from the original work. The instability persisted both from the original sequence length setting of 20 and our customized setting of 100. The main obstacle came from the discriminator vastly outperforming the generator. Even after pretraining the generator to achieve a saturated performance, the generator failed to fool the discriminator, and the discriminator identified all the given sequences as fake with extremely high confidence (close to 1), which provided no meaningful reward signals.

We thus lowered the representational power of the discriminator by reducing the number of 1-D convolutional layers from 10 to 5. We also increased the receptive field of convolution filters up to 20 (and discarded layers with small size filters), since we wanted the discriminator to capture a periodic structure of musical sequences effectively. Note that the large receptive field approach is shown to be effective in the related work, which handles raw waveform audio \cite{donahue2018synthesizing}. 

Furthermore, we found that hyperparameters for policy gradients needed careful optimization. We used 32 (instead of 16) Monte Carlo search rollouts for calculating rewards in the policy gradient to ensure lower variance of reward signals. This prevented the generator from learning with an unnecessary noise, which would lead to divergence and critically impact performance of the model. We adjusted the reward discount factor from 0.95 to 0.99 to compensate for the longer sequence length of 100. We also applied a more ``conservative'' target generator network update rate from 0.8 to 0.9. We observed that the higher update rate (i.e. less amount of parameter update of the target network) stabilized the adversarial training with reward signals and constrained the divergence of the generator.

% GANhack 방법 설명
Instead of feeding a mixture minibatch containing both real and fake samples to the discriminator as in the original work, we used minibatch discrimination technique where minibatches contained only real or fake samples. This technique is used in several other works with GANs \cite{metz2016unrolled}, and it empirically improved adversarial training of the model.

\section{Experimental Results}
% 확장된 SeqGAN 모델 사이즈, hyperparameters, training scheme 서술
We trained SeqGAN with hyperparameter optimization, which resulted in a larger version of the original model. Our polyphonic word representation of a MIDI file has a vocabulary size of 3,216. We embedded each word with randomly initialized 32-dimensional vectors. We created sequences of length of 100 for training. This length also applies to sequence generation from the trained model. The generator RNNs have 512 LSTM cells. The discriminator CNNs have five 1-D convolutional layers, and each of them has 400 feature maps with a receptive field of 20. We pretrained both generator RNNs and discriminator CNNs for 100 epochs with the regular negative log-likelihood loss. Due to the tendency of the discriminator to dominate, we first pretrained the generator and the discriminator at learning rates of 0.001 and 0.0001 and set the learning rate of the generator higher at 0.01. We used a batch size of 32 for all experiments.

% sample generation 방법 및 성능 측정법 서술: training & validation NLL, BLEU score
We compared two strategies: the unconditional method where sampled sequences always started from the pre-defined zero token, and the conditional method where we trained the model and generated sequences from the trained model with the first word in the real sequence as a start token. For each strategy, we additionally compared two formulations of the loss for the discriminator: the original softmax reward with the cross entropy loss and a sigmoid reward with the least squares loss, which is known to stabilize the training of GANs \cite{mao2017least}. The generator followed the same policy gradient method with the given scalar reward in each time step. The generated sequences showed musically coherent structure with long-term harmonics. We measured results both from quantitative and perceptive qualitative perspectives.

For quantitative analysis, we calculated the BLEU score that measures a similarity between the validation set and the generated samples and which is largely used to evaluate the quality of machine translation \cite{papineni2002bleu}. To be specific, the BLEU score can be calculated by comparing the entire corpus from the validation set and the sequence generated from the model. A higher BLEU score means that samples from the generator follow the underlying real data distribution more closely. For the conditional method, we used a start token from a randomly sampled batch from the training set.

\begin{table}
 \begin{center}
 \begin{tabularx}{\columnwidth}{c|c|c}
  Algorithm & Log-likelihood & Adversarial\\
  \hline
  SeqGAN, Uncond.  & 0.5335 & 0.6272\\
  \hline
  SeqGAN, Cond.  & 0.5095 & 0.5552\\
    \hline
    \hline
  LS-SeqGAN, Uncond.  & 0.5312 & \textbf{0.6852}\\
    \hline
  LS-SeqGAN, Cond.  & 0.5177 & 0.5743\\
 \end{tabularx}
\end{center}
 \caption{Performance comparison with BLEU-4 scores from the validation set. SeqGAN: original softmax output from the discriminator with the cross entropy loss. LS-SeqGAN: sigmoid output from the discriminator with the least squares loss.}
 \label{bleuscore}
\end{table}

\begin{table}
 \begin{center}
 \begin{tabular}{c|c|c|c}
  Sample & Pleasant? & Real? & Interesting?\\
  \hline
  Uniform Random & 2.36 & 2.10 & 2.71 \\
  \hline
  \hline
  Log-likelihood & 3.02 & 2.93 & 3.07 \\
  \hline
  Adversarial 1 & 3.17 & 2.69 & 3.14 \\
  \hline
  Adversarial 2 & \textbf{3.90} & 3.62 & \textbf{3.83} \\
  \hline
  Adversarial 3 & 3.86 & \textbf{3.81} & 3.76 \\
  \hline
  Mode Collapse & 1.67 & 1.67 & 1.90 \\
  \hline
  \hline
  Real Sample & \textbf{4.31} & \textbf{4.31} & \textbf{4.07} \\
 \end{tabular}
\end{center}
 \caption{Mean Opinion Score (MOS) results. Uniform Random: a sample generated with a uniform random probability in each time step from the vocabulary. Adversarial: samples from adversarial training with progressively increasing BLEU-4 score. Mode Collapse: a sample from the failure case of adversarial training with BLEU-4 score below 0.2.}
 \label{mostable}
\end{table}

\tabref{bleuscore} showed that the BLEU score of the generator RNN is saturated from the pretraining and is further improved by the adversarial training. The generator RNN trained with NLL loss showed peak performance when the BLEU score reached approximately 0.53 and the adversarial training could generally improve the score from 0.05 to 0.15. The best configuration had the BLEU score of over 0.68. Note that these improvements are similar in magnitude to those reported in the original paper. However, we could not reproduce the same results with the original network configurations because of the instant divergence of the generator.

Results showed that the unconditional method performed relatively better than the conditional method especially in the adversarial training phase. A possible explanation is that the unconditional method can estimate manifolds from the embedded space better with the fixed zero start token, because the model can observe many more trajectories of the real data manifold from a single starting point, compared to smaller number of trajectories from many starting points in the pretraining phase of the conditional method. This further impacts potential benefits from the unsupervised adversarial training with reinforcement learning signals, as the model pretrained with the conditional method tends to fall into a bad local minimum with a higher probability than the model pretrained with the unconditional method.

We conducted a qualitative analysis of human perceptive performance of the generated MIDI sequences using MOS user study. The experiment asked 42 participants to rate seven different sequences from 1 to 5, by responding to three questions: How pleasant is the song? How realistic is the sequence? How interesting is the song? These questions are constructed given the inspiration from MidiNet. The seven sequences included a sample from a real dataset, a sequence sampled by uniform random probability in each time step from the vocabulary, and a sample from a failure case of the adversarial training with low BLEU score (below 0.2). To remove the bias, we notified participants that all seven sequences were generated by the model. \tabref{mostable} showed that the sequences from adversarial training sounded more like the real ones than the sequences from the pretrained model with NLL, which is consistent with the quantitative analysis.

Samples from the model pretrained with NLL sounded relatively more repetitive and focused more on the short-term harmonics. This is to be expected, since the pretraining phase targets the next token in the real training dataset. Samples from the adversarial training tended to show longer harmonics with more consistent chord progressions, possibly since the model successfully explored policies that received high reward by keeping the chord progression.

\section{Discussion and Future Work}
% GAN이 대부분 불안정하고 BLEU 오히려 떨어질때도 많으며, mode collapsing 발생하고 NLL에 비해 가성비가 안좋다는 점 서술?
Although experiments showed that the adversarial training further boosted the performance in the music language modeling, there are drawbacks due to the nature of GANs. Firstly, GANs often suffer from the mode collapsing problem, where the generator fools the discriminator by creating artifacts rather than realistic samples \cite{metz2016unrolled}. We also have noticed this problem where the generated samples played the same note constantly, which broke the musical coherence. This phenomenon can also be observed with a decrease in the BLEU score, which implies a divergence from the pretrained model. Recent works on GANs introduce earth-mover distance as a loss function to overcome this issue \cite{arjovsky2017wasserstein}. Thus, incorporating this idea to discrete GANs could alleviate the problem \cite{kim2017adversarially}. There have been recent improvements in the original work based on the rank-based loss \cite{lin2017adversarial}, which can be directly applicable to our task.

Secondly, the training of GANs is not more computationally efficient than the NLL training of the generator RNNs. For example, with our stabilized hyperparameters, GANs require roughly ten times more computing time than the NLL training per epoch for a relatively small improvement in performance. The computational cost also scales to the number of Monte Carlo policy rollouts, which gives us a trade-off between accuracy and variance. 

Thirdly, the policy gradient method with the Monte Carlo rollout is highly stochastic. Although the adversarial training can provide the extra performance improvement that the NLL method cannot, the reinforcement learning signal showed high variance and a relatively low reproducibility. This means that even for the same hyperparameter settings, one would need to run multiple training trials to achieve improvements from the adversarial training. This leaves room for improvements in minimizing the variance of the reinforcement learning signals notably by Monte Carlo Tree Search (MCTS) \cite{browne2012survey} and experience replay \cite{mnih2015human} as examples.

% hashing 의 단점 서술 -> unconstrained 모델 구조 필요
The restriction of the vocabulary to the pre-defined words that are observed in the dataset has a limitation that the model cannot create chords and melodies that are outside the dataset. In terms of creativity, the model would have to ``compose'' a novel music outside the boundaries of the learned data \cite{briot2017deep}. While harder to train, the unconstrained models capable of processing arbitrary polyphonic input and output are crucial for creativity.
%% unconstrained 모델을 위해 RL signal을 이용할 계획이라 이걸 가장 주요 문제로 삼고 맨 앞으로 배치하였습니다

% RL signal 활용은 music theory 가 이미 존재하기 때문에 특히 중요
%% 앞에 한문장 추가했습니다.
As we have mentioned in the related work, we have observed that reinforcement learning using reward signals is a direct way to inject prior knowledge about musical structure into the model. This suggests that we could further leverage the reinforcement learning signals by incorporating a critic model that evaluates musical consonance based on music theory. Indeed, RL-Tuner, a deep Q-networks based model, uses scores from music theory rules as auxiliary reward signals \cite{jaques2017tuning}. We plan to implement this idea in the future work.

% word embedding 대신 CNN 활용
Albeit the proposed word embedding method for the polyphonic MIDI data is simple and efficient, the word embedding with random projection does not effectively capture relative harmony and consonance of each word. Modular networks that consider this relative information of the MIDI data could further improve performance of the music language model. CNNs are a viable choice for this purpose \cite{kim2014convolutional}, and we plan to use the CNN-RNN hybrid model in the future work. 

For more objective and structured experiments with automatic music generation, we need a robust quantitative measures to evaluate the perceptive quality of the machine-generated music \cite{briot2017deep}. From our experiments, the quantitative BLEU score analysis was consistent with the qualitative MOS user study to a certain degree, but did not exactly reflect the perceptive performance. Development of a structured quantitative metric would improve objectivity and reproducibility of research on the automatic music generation.

% For bibtex users:
\bibliography{main}

% For non bibtex users:
%\begin{thebibliography}{citations}
%
%\bibitem {Author:00}
%E. Author.
%``The Title of the Conference Paper,''
%{\it Proceedings of the International Symposium
%on Music Information Retrieval}, pp.~000--111, 2000.
%
%\bibitem{Someone:10}
%A. Someone, B. Someone, and C. Someone.
%``The Title of the Journal Paper,''
%{\it Journal of New Music Research},
%Vol.~A, No.~B, pp.~111--222, 2010.
%
%\bibitem{Someone:04} X. Someone and Y. Someone. {\it Title of the Book},
%    Editorial Acme, Porto, 2012.
%
%\end{thebibliography}

\end{document}